\documentclass[twocolumn]{aastex701}
\usepackage{natbib}
\usepackage{subcaption}
\usepackage{graphicx}
\usepackage{amsmath}
\begin{document}


\title{Compact Ca\,\textsc{II} K Brightenings Precede Solar Flares: A Dunn Solar Telescope Pilot Study}

\accepted{for publication in ApJL on December 17, 2025}

\author[orcid=0000-0001-6673-6000,sname='A Kumar']{Aman Priyadarshi M. Kumar}
\affiliation{Department of Astronomy, New Mexico State University, Las Cruces, NM 88003, USA}
\email{kumar@nmsu.edu}

\author[orcid=0000-0002-4188-7010,sname='J Shetye']{Juie Shetye}
\affiliation{Department of Astronomy, New Mexico State University, Las Cruces, NM 88003, USA}
\email{jshetye@nmsu.edu}  

\author[orcid=0000-0001-5342-0701,sname='S G Sellers']{Sean G Sellers}
\affiliation{Department of Astronomy, New Mexico State University, Las Cruces, NM 88003, USA}
\email{sellers@nmsu.edu}  

\author[0000-0003-1746-3020]{Damian J. Christian}
\affiliation{ Department of Physics and Astronomy, 
California State University Northridge, 
18111 Nordhoff St, Northridge, CA 91330, USA }
\email{damian.christian@csun.edu}

\begin{abstract}

We present a uniform analysis of compact Ca\,\textsc{ii}  K (3934\,\AA) brightenings that occur near flare kernels and assess their value as short-lead indicators of solar flare onset. Using high-cadence imaging from the Rapid Oscillations in the Solar Atmosphere (ROSA) instrument at the Dunn Solar Telescope (DST), we examine eight flare sequences (seven C-class and one B-class) obtained between 2021 and 2025. Fixed, detector-coordinate regions of interest are used to generate mean-intensity light curves, which are detrended and smoothed to isolate impulsive brightenings. In every event, a compact Ca\,\textsc{ii}  K brightening is detected within or adjacent to the flaring region that peaks 10–45\,min before the primary kernel and the corresponding rise in GOES 1–8\,\AA\ flux. The measured temporal offsets scale with the deprojected separation between the brightening and flare kernels, implying an apparent propagation speed of $\sim$30–35\,km\,s$^{-1}$ that is consistent with chromospheric reconnection. Complementary Spectropolarimeter for Infrared and Optical Regions (SPINOR) spectropolarimetry for one event shows topological reconfiguration from closed to open or extended connectivity, supporting a reconnection-driven origin. These results demonstrate that compact Ca\,\textsc{ii}\,K brightenings are reproducible, physically meaningful precursors to flare onset. Their simplicity and cadence make them attractive chromospheric indicators, and future work will evaluate their predictive skill alongside established UV/EUV and magnetic diagnostics.

\end{abstract}



\section{Introduction} 

Solar flares are impulsive releases of magnetic free energy in the solar atmosphere (up to $\sim10^{32}$ erg), produced when stressed coronal fields reconnect and rapidly convert magnetic energy into heating, radiation, mass motions, and particle acceleration \citep[e.g.,][]{Priest2002,ShibataMagara2011,Benz2017}. Flare energy is stored by shear/twist and flux rope formation above polarity-inversion lines \citep[e.g.,][]{vanBallegooijen1989}, while eruptions are then triggered by magnetic reconnection and/or ideal MHD instabilities in several well-studied scenarios \citep[e.g.,][]{Canfield1999}, including tether-cutting reconnection in sheared arcades \citep[e.g.,][]{Moore2001}, breakout reconnection in multipolar topologies \citep{Antiochos1999}, and the kink/torus instabilities acting on pre-existing or newly built flux ropes \citep{hoodpriest1979,TorokKliem2005}. 

Given the operational emphasis on imminent onsets, a parallel literature targets pre-flare signatures minutes to tens of minutes before the impulsive phase, compact chromospheric/transition-region brightenings and line broadening, small jets near polarity inversion lines (PILs), and slow-rise filaments or hot channels  \citep[e.g.,][]{Chifor2007,Bamba2013}. These precursors appear where low-altitude reconnection perturbs a highly sheared arcade or nascent rope, consistent with physics-based trigger models \citep{Kusano2012,Bamba2017}. 

For scale, flare occurrence statistics indicate that small events dominate the solar flare population. A recent GOES/XRS analysis of 18{,}833 events finds a peak-irradiance distribution with slope $\alpha_{\mathrm{peak\,irrad}}\approx2.16\pm0.01$, with B- and C-class flares vastly outnumbering M/X events \citep[][their Fig.~1]{Mason2023}. This steep distribution argues for lightweight, high-throughput indicators capable of monitoring frequent, low-energy activity from the ground.

Beyond the Sun, hydrogen Ly$\alpha$ is the dominant FUV line of cool stars and shows strong intrinsic emission and variability: HST programs have reconstructed or directly measured K/M-dwarf Ly$\alpha$ profiles and variability \citep{France2013,Youngblood2016,Peacock2025}, while time-resolved FUV campaigns reveal flare-driven Ly$\alpha$ excursions and tens-of-percent variability that can mask exospheric signals \citep{Loyd2018,Rockcliffe2025}. Reviews of stellar flares emphasize Ly$\alpha$ and Balmer diagnostics as sensitive responders to impulsive heating and nonthermal particles \citep{Kowalski2024LRSP}. Foundational stellar work links Ly$\alpha$ emission with activity and winds via astrospheric absorption \citep{Linsky2014}. For the Sun, Ly$\alpha$ brightens promptly during flares and often tracks thermal soft X-rays \citep{Milligan2020}; proxy and reconstruction efforts include early solar Ly$\alpha$ proxies from He \textsc{i} 10830 \AA\, \citep{Li2022}. This broader context motivates our chromospheric study: we target short-lead, ground-accessible precursors and show that Ca\,\textsc{ii}  K compact brightenings provide operationally simple triggers complementary to UV/EUV spectroscopic methods.

Ground-based high-resolution imaging spectroscopy now provides complementary, multi-wavelength constraints on that vertical coupling and on flare onset/evolution. With the Swedish 1-m Solar Telescope \citep[SST,][]{Scharmer2019SST}, the CRisp Imaging SpectroPolarimeter  \citep[CRISP;][]{Scharmer2008CRISP} and the CHROMOspheric Imaging Spectrometer \citep[CHROMIS;][]{Lofdahl2021SSTRED} have resolved ribbon substructure at $\sim$100–200 km scales and periodic “blobs” consistent with fragmented reconnection/tearing in the current sheet, together with line-profile asymmetries that diagnose flows across the chromosphere \citep{Faber2025,Pietrow2024}. SST Ca\,\textsc{ii}  8542\AA\, inversions further show strong footpoint heating and flare-driven chromospheric condensations during the rise/peak phases, while H$\alpha$/Ca\,\textsc{ii}  K kinematics capture fast ribbon-parallel motions and quasi-periodic bursts \citep{Kuridze2017,Keys2011}. Using the Dunn Solar Telescope (DST), Interferometric BIdimensional Spectrometer (IBIS) spectropolarimetry has revealed stepwise chromospheric magnetic-field changes during major flares, spatially tied to loop footpoints and distinct from photospheric signatures—direct evidence for rapid coronal-to-chromospheric coupling \citep{Kleint2017,Zbinden2024}. Early Daniel K. Inouye Solar Telescope (DKIST) campaigns add ultra-high-resolution H$\alpha$ views of flare arcades and exquisitely structured ribbons, opening sub-50 km scales in the decay phase and during coordinated space–ground runs \citep{DKISTVBI2025,DKISTCoord2025}. While space-borne spectra remain powerful, these results show that modern ground facilities deliver fast-cadence, fine-scale diagnostics (H$\alpha$, Ca\,\textsc{ii}  K/8542, and spectropolarimetry) well suited to fast-cadence studies of flare onset, which we pursue here in Ca\,\textsc{ii}\,K.

Motivated by these ground-based results, we evaluate a complementary, operationally simple approach using Ca\,\textsc{ii}~K (3934\,\AA) imaging at the Dunn Solar Telescope (DST). The physics is straightforward: early, low-altitude energy deposition heats and ionizes chromospheric plasma, producing compact Ca\,\textsc{ii}  K kernels near future ribbon sites, whereas soft X–ray (SXR) emission peaks later as hot coronal plasma accumulates and the emission measure grows. We use high-cadence Rapid Oscillations in the Solar Atmosphere \citep[ROSA,][]{Jess2010} Ca\,\textsc{ii}  K sequences from the Sunspot Solar Observatory Data Archive \citep[SSODA;][]{Sellers2025}.

We provide a practical recipe and uncertainty handling, and we document guardrails for ambiguous cases (closely spaced sub–flares, partial temporal coverage, or stationary plage). Section~\ref{sec:data} details the data and regions of interest (ROI) pipeline; Section~\ref{sec:methodsandresult} compiles event overlays demonstrating the precursor brightening; Section~\ref{sec:conclusion} addresses limitations and integration with existing systems.

\section{Data} \label{sec:data}

We analyzed chromospheric Ca\,\textsc{ii} K  (3934\,\AA) sequences obtained with the Rapid Oscillations in the Solar Atmosphere (ROSA) instrument at the DST in Sunspot, New Mexico. Fully speckle-reconstructed and co-aligned data products were accessed from the Sunspot Solar Observatory Data Archive \citep[SSODA;][]{Sellers2025}, which provides calibrated ROSA datasets with accompanying World Coordinate System (WCS) headers for geometric consistency. In the configuration used here, the Ca\,\textsc{ii}~K channel employs a narrowband interference filter centered at 3934\,\AA\ with FWHM $\approx1.2$\,\AA\ on a $1002\times1004$ detector.  The plate scale values recorded in the SSODA headers for our sample span
$0\farcs06$--$0\farcs16$\,pix$^{-1}$, corresponding to fields of view of
roughly $60^{\prime\prime}$--$160^{\prime\prime}$ on a side. SSODA Level~1 ROSA products are generated in ``burst mode'', in which $N_{\rm burst}$ short-exposure images are dark- and flat-corrected, grouped into bursts, and reconstructed with the \textsc{KISIP} speckle algorithm \citep{kisip2}; in our Ca\,\textsc{ii}~K sequences $N_{\rm burst}=64$ for all events except 2022 April 18, which uses $N_{\rm burst}=32$. All basic preprocessing steps such as dark and flat correction, speckle reconstruction, and image alignment follow the standard SSODA pipeline described by \citet{Sellers2025}.

H$\alpha$ image sequences were recorded with Hydrogen-Alpha Rapid Dynamics camera (HARDcam), a high-cadence narrowband 6563 \AA{} imager co-mounted with ROSA at the DST.  {The HARDcam channel uses a Lyot filter centered at 6562.8\,\AA\ with FWHM $\approx0.25$\,\AA\ and a $2048\times2048$ detector, giving a plate scale of $0\farcs0845$\,pix$^{-1}$ and a field of view of $\approx173^{\prime\prime}\times173^{\prime\prime}$.  Native frame rates are $\approx29$\,Hz, and the Level~1 H$\alpha$ products distributed by SSODA are likewise produced from bursts of 64 short-exposure frames reconstructed with \textsc{KISIP} \citep{kisip2}.} HARDcam provides sub-second sampling and arcsecond-scale resolution over the ROSA field, and processed products are distributed via SSODA with WCS metadata for co-registration \citep{Jess2010,Sellers2025}.

The analyzed sample includes eight DST flare sequences obtained between 2021 and 2025. These events were selected for their clear chromospheric activity in Ca\,\textsc{ii}  K and concurrent GOES 1--8\,\AA\ coverage.  {Across the sample, the effective Level~1 cadences for the Ca\,\textsc{ii}~K and H$\alpha$ sequences are a few seconds, set by the ratio of the native camera frame rates (6--30\,Hz for ROSA Ca\,\textsc{ii}~K and $\approx29.4$\,Hz for HARDcam H$\alpha$) to the burst sizes quoted above.  A summary of the key Ca\,\textsc{ii}~K parameters (plate scale, field of view, and Level~1 cadence) for each event is given in Table~\ref{tab:timing}, and the modest variations between runs do not affect the relative timing or intensity measurements presented below.}

\begin{figure*}[ht!]
\centering
\includegraphics[width=\linewidth]{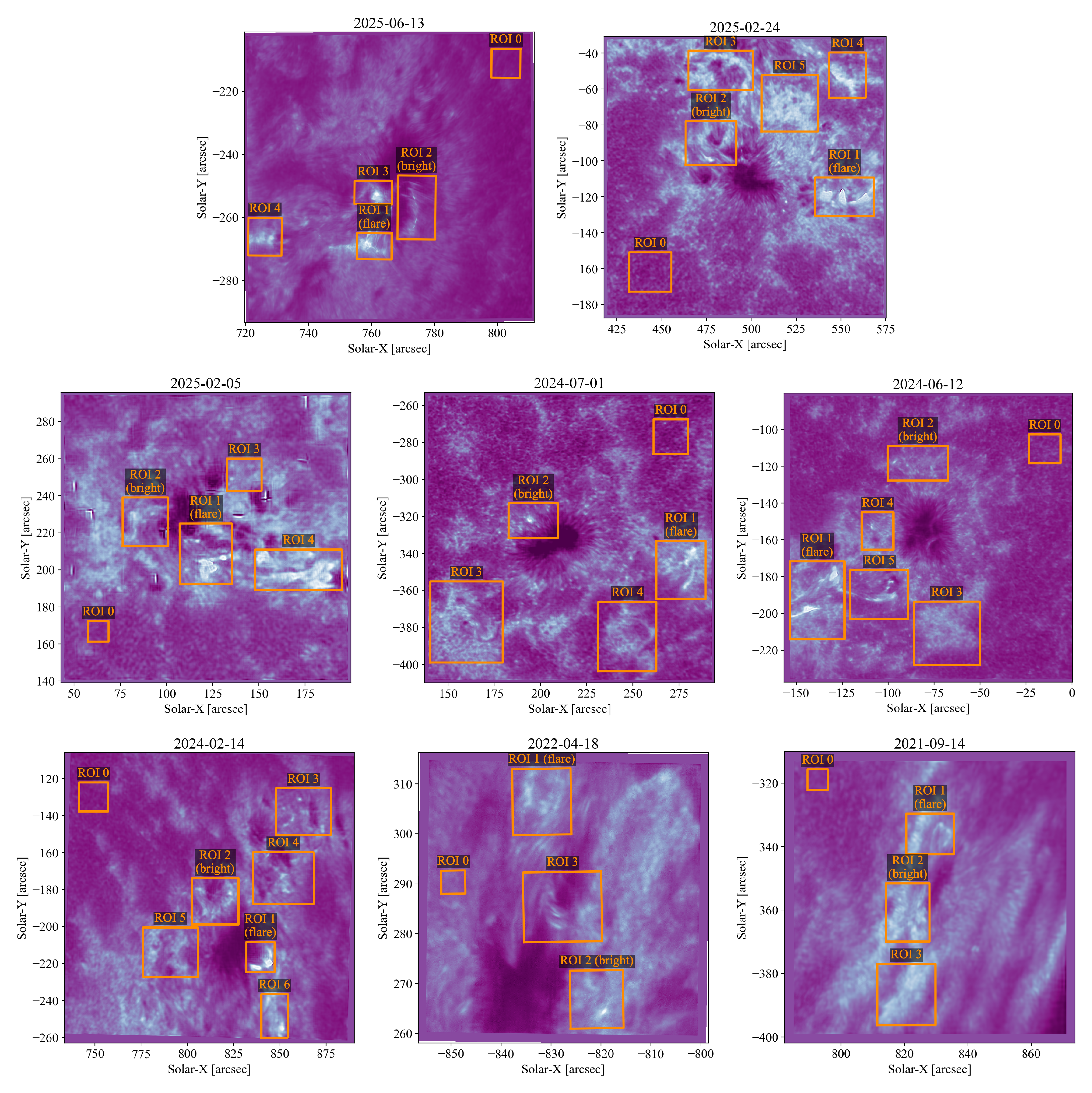}
\caption{
Context Ca\,\textsc{ii}  K images (ROSA/DST) for the eight analyzed events. 
Orange rectangles mark fixed ROIs.  {The ROI labels are standardized across all events: ROI~1 marks the flare kernel, ROI~2 the precursor brightening, ROI~0 the quiescent reference region, and ROIs~3 and higher denote additional contextual locations.}
}
\label{fig:roi_map}
\end{figure*}

For each event, a compact set of fixed rectangular 
ROIs were defined to sample the primary flaring kernel, nearby compact brightenings, and a quiet reference area within the ROSA field. Figure~\ref{fig:roi_map} shows the Ca\,\textsc{ii}  K context images for all eight events, with ROIs marked in orange; {for consistency across all events we adopt a fixed labeling scheme in which ROI~1 always denotes the principal flare kernel, ROI~2 the compact brightening analyzed as the precursor, ROI~0 a quiescent reference region used to track background and instrumental fluctuations, and ROIs~3 and higher denote additional contextual locations that do not enter the timing statistics.} The basic timing properties of these events including the start, peak, and end times of the primary Ca\,\textsc{ii}  K kernels (flaring regions) are summarized in Table~\ref{tab:timing}. These times define the reference flare peaks used in subsequent analysis. 

\begin{deluxetable*}{lcccccccccc}
\tabletypesize{\scriptsize}
\tablewidth{\textwidth}
\tablecaption{ {Per–event Ca\,\textsc{ii}  K timings for the standardized flare ROI (ROI~1), basic ROSA Ca\,\textsc{ii}~K image geometry, and Level~1 cadence.}\label{tab:timing}}
\tablehead{
\colhead{Date} &
\colhead{GOES} &
\colhead{Ca\,\textsc{ii} K Start} &
\colhead{Ca\,\textsc{ii} K Peak} &
\colhead{Ca\,\textsc{ii} K End} &
\colhead{Impulsive} &
\colhead{Decay} &
\colhead{Total} &
\colhead{ {$s_{\rm CaK}$}} &
\colhead{ {FOV$_{\rm CaK}$}} &
\colhead{ {Cadence$_{\rm CaK}$}} \\
\colhead{(UTC)} &
\colhead{Class} &
\colhead{(UTC)} &
\colhead{(UTC)} &
\colhead{(UTC)} &
\colhead{(min)} &
\colhead{(min)} &
\colhead{(min)} &
\colhead{ {($^{\prime\prime}$ pix$^{-1}$)}} &
\colhead{ {($^{\prime\prime}\!\times\!^{\prime\prime}$)}} &
\colhead{ {(s)}}}
\startdata
2025/06/13 & C3.1 & 13:45:35 & 13:57:25 & 14:18:02 & 11.83 & 20.63 & 32.45 &  {0.093}  &  {$93\times93$}   &  {4.22}  \\
2025/02/24 & C3.7 & 15:27:03 & 15:40:13 & 16:04:30 & 13.16 & 24.29 & 37.45 &  {0.156} &  {$157\times156$} &  {4.22}  \\
2025/02/05 & C4.6 & 16:13:26 & 16:22:10 & 17:01:06 & 8.73  & 38.93 & 47.66 &  {0.156} &  {$157\times156$} &  {4.22}  \\
2024/07/01 & C4.0 & 13:48:21 & 14:00:53 & 14:18:59 & 12.53 & 18.09 & 30.62 &  {0.156} &  {$157\times156$} &  {4.22}  \\
2024/06/12 & C3.3 & 15:15:45 & 15:21:48 & 15:35:36 & 6.05  & 13.80 & 19.85 &  {0.156} &  {$157\times156$} &  {4.22}  \\
2024/02/14 & C3.0 & 16:03:40 & 16:06:21 & 16:26:37 & 2.68  & 20.28 & 22.95 &  {0.156} &  {$157\times156$} &  {4.22}  \\
2022/04/18 & C3.9 & 14:11:07 & 14:21:50 & 14:43:58 & 10.72 & 22.12 & 32.84 &  {0.059} &  {$59\times59$}   &  {3.17}  \\
2021/09/14 & B2.0 & 16:33:08 & 16:39:17 & 17:06:03 & 6.16  & 26.75 & 32.91 &  {0.092} &  {$92\times92$}   &  {10.56} \\
\enddata
\end{deluxetable*}

\begin{figure*}[ht!]
    \centering
    \includegraphics[width=0.83\textwidth]{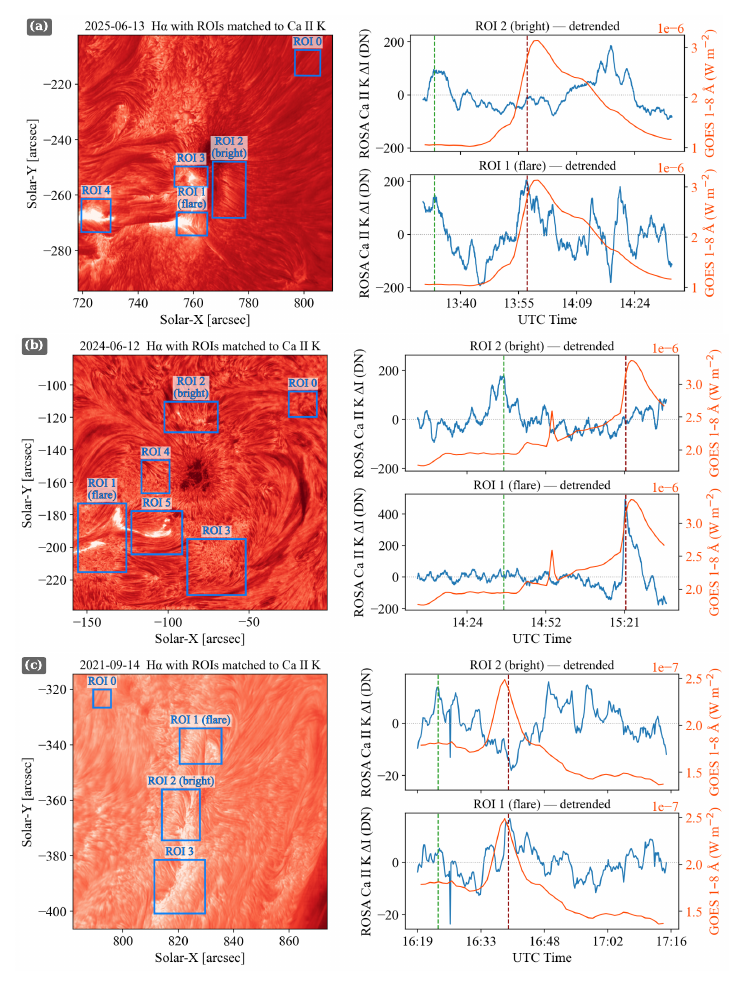}
    \caption{Multi--wavelength timing of precursor and flare kernels.
    For each representative event, the left panel shows the H$\alpha$ context image from DST/HARDcam near the time of onset, the right upper panel shows the detrended Ca\,\textsc{ii}  K light curve of the compact brightening site, and the right lower panel shows the corresponding flare kernel.
    In both light curves, dashed green and red lines mark the brightening and flare peak times, respectively, while the vermilion curve shows the co-temporal GOES 1--8\,\AA\ soft X-ray flux (right $y$-axis).
    The temporal ordering shows that the compact Ca\,\textsc{ii}\,K brightening leads the main flare kernel and the soft X-ray rise, identifying it as the earliest detectable chromospheric energy release in these events.
    Panels (a)--(c) correspond to the events on 2025-06-13, 2024-06-12, and 2021-09-14.
     {In all panels ROI~1 labels the flare kernel, ROI~2 the precursor brightening, and ROI~0 a quiescent reference patch; this convention matches the Ca\,\textsc{ii}\,K ROI maps.}
}
    \label{fig:bright_flare_goes}
\end{figure*}

The combination of high cadence, stable pointing, and fixed ROIs enables each region to produce a clean mean-intensity light curve suitable for automated detrending and peak extraction. Because the ROIs are defined directly on the reconstructed frames, no additional segmentation or tracking is required, making the procedure reproducible and computationally efficient.

In addition, for one event, we present high-resolution spectropolarimetry of the active region acquired with the SPectropolarimeter for INfrared and Optical Regions \citep[SPINOR,][]{Socas-Navarro2006} at the DST. Several of the events in the larger catalog are complemented by SPINOR observations, from which we have selected SOL2025-06-13 for further study. During this observation sequence, SPINOR was operated with two cameras simultaneously observing the Fe~\textsc{i}~5250~\AA\ and Ca\,\textsc{ii} ~8542~\AA\ lines, which provide magnetic sensitivity in the low photosphere and low chromosphere respectively. These lines provide detailed diagnostics of the magnetic and thermodynamic structure of the active region. Full-4D (two spatial axes, one spectral, and one polarimetric) maps of the active region were acquired by scanning the spectrograph slit across the sunspot in a dense-rastering fashion. At each slit position, the full Stokes vector was observed for the full wavelength range along the slit. Over the $\approx26.6$~minute raster duration, the instrument covered an area of $104^{\prime\prime}\times87^{\prime\prime}$. From these data, we derive maps of magnetic connectivity in the region, providing further insight into the regions highlighted as Ca\,\textsc{ii}  K precursor sites. These data can be found in Section~\ref{sec:spinor_maps}.


\section{Methods and Results} \label{sec:methodsandresult}

While all eight sequences were analyzed, three representative cases (2025 June 13, 2024 June 12, and 2021 September 14) are highlighted in Figure~\ref{fig:bright_flare_goes} to illustrate, in H$\alpha$, the typical morphology and evolution.

We analyzed each Ca\,\textsc{ii} K sequence using multiple regions of interest (ROIs) that sample the primary flare kernel, nearby compact brightenings, and quiet reference areas within the same field of view. We use the standardized ROI labeling scheme defined in Section~\ref{sec:data}. For every ROI, the mean intensity was measured in each frame to produce a uniformly sampled light curve at the native Level-1 cadence of a few seconds (see Table~\ref{tab:timing}). From these curves, the site showing the earliest impulsive rise relative to the main kernel was identified as the precursor brightening.

 {In several events (notably 2024 June 12, 2025 April 18, and 2021 September 14) the Ca\,\textsc{ii}~K light curves in and around the flare site exhibit a sequence of peaks both before and after the flare, and more than one ROI shows impulsive brightenings. In such cases the association between a single compact brightening and the eventual flare kernel is inherently ambiguous: the same active region dynamics that produce the flare also drive background variability in nearby plage and ribbon fragments. For this pilot study we select, by visual inspection, the ROI whose morphology and timing most clearly link it to the flare, and we treat the resulting $\Delta t$ as illustrative rather than as a unique, fully automatic flare-onset proxy for that event. A more stringent operational use of Ca\,\textsc{ii}~K brightenings will require additional morphological and magnetic constraints to down–select among multiple candidate peaks, as discussed further in Appendix~\ref{sec:appendixA}.}


To isolate the impulsive structure, each ROI light curve is detrended by subtracting a low–order (quadratic) background and is lightly smoothed over a few frames. This step suppresses slow transparency or seeing variations while preserving the sub-minute timing (explained in greater detail in Section~\ref{sec:appendixA} of the appendix). The resulting flattened series cleanly reveal impulsive brightenings and allow direct temporal comparison among regions. For each ROI we determine the start, peak, and end times of the dominant brightening using the zero–crossings of the first derivative, requiring a prominence greater than $3\sigma$ relative to a running-median background. Peaks within 60\,s are merged to avoid overcounting. The temporal offset between the brightening and flare kernels is defined as
\begin{equation}
  \Delta t = t_{\mathrm{peak}}^{(\mathrm{flare})} - t_{\mathrm{peak}}^{(\mathrm{bright})},
\end{equation}
so that $\Delta t>0$ means the compact brightening peaks earlier than the main flare kernel. Typical timing errors are $\sigma_{\Delta t}\!<\!0.5$\,min which are negligible compared with the multi-minute offsets observed across the ensemble.

 {It is important to note that some methodological caveats apply to this timing procedure. Partial temporal coverage can bias peak finding if the rise or decay phase is truncated. Closely spaced sub-flares may produce competing maxima, although our 60\,s merging window and $>3\sigma$ prominence threshold reduce such ambiguities. The GOES 1--8\,\AA\ light curve is quantized to 1\,min cadence, introducing a small discretization relative to the ROSA cadence. Finally, the present pipeline is purely timing-based and does not yet incorporate explicit intensity thresholds or ribbon-evolution constraints; these limitations mainly add modest random or systematic errors to individual events without altering the multi-minute lead times seen across the sample.}

\subsection{Geometry, temporal offset, and foreshortening}

For each event we measure the projected center–to–center separation between the precursor and flare ROIs using the image WCS. Apparent distances shrink toward the limb, so we correct for foreshortening using the heliocentric angle $\theta_{\rm mid}$ at the ROI midpoint:
\begin{equation}
  D = \frac{d_{\mathrm{proj}}}{\mu}, \qquad \mu = \cos\theta_{\mathrm{mid}},
\end{equation}
yielding a deprojected physical separation $D$ in megameters. Uncertainties combine plate-scale calibration, ROI centroiding, and geometric projection terms from the finite ROI size.

Each Ca\,\textsc{ii} K sequence is aligned with the corresponding GOES 1--8\,\AA\ light curve to compare chromospheric and coronal timing. Figure~\ref{fig:bright_flare_goes} shows three representative examples (2025 June 13, 2024 June 12, and 2021 September 14). In all cases the Ca\,\textsc{ii} K brightening rises and peaks before the flare kernel and the onset of the GOES soft X-ray increase, confirming that the compact brightening represents the earliest detectable energy release in the system. The measured temporal offsets, typically 10--45\,min with a mean of $\sim$27\,min, are far larger than any instrumental latency, establishing a genuine physical lead. A more quantitative examination of how $\Delta t$ scales with the deprojected distance $D$, including linear fits and correlation coefficients, is presented in Appendix~\ref{sec:appendixB} (Figure~\ref{fig:geom_timing_speed}); here we simply note that any distance–time trend is modest and serves as supporting context rather than a central result.

\subsection{Physical interpretation}

 {The combined geometry and timing are consistent with a picture in which compact Ca\,\textsc{ii}  K brightenings mark localized, low-altitude energy releases that precede the flare kernel by several minutes. In an ensemble sense there is only a weak trend for $\Delta t$ to increase with deprojected distance $D$ (Appendix~\ref{sec:appendixB}); if one fits a straight line, the slope can be expressed as an apparent propagation speed of order 30--35\,km\,s$^{-1}$. However, the corresponding Pearson and Spearman coefficients are modest and rely on linear or monotonic relationships that may not capture more complex behavior, so we regard this trend as suggestive rather than definitive and use the implied speed only as an order-of-magnitude guide.} These findings align with reports of small transient brightenings near polarity-inversion lines that herald large-scale ribbon activation \citep[e.g.,][]{Chifor2007,Bamba2013}. The simplicity and ground accessibility of the Ca\,\textsc{ii}  K signal, coupled with its fast cadence, highlight its operational potential as a real-time flare onset indicator.

Overall, the ensemble supports a physically consistent sequence: an early chromospheric brightening appears minutes before the main flare, with temporal offsets scaling with the physical separation and viewing geometry. This behavior suggests that precursor emission traces an expanding reconnection or heating front advancing toward the flare kernel at tens of km\,s$^{-1}$.

\subsection{Example magnetic configuration traced with SPINOR}\label{sec:spinor_maps}

\begin{figure*}[t]
  \centering

  \begin{subfigure}[t]{0.82\textwidth}
    \centering
    \includegraphics[width=\linewidth]{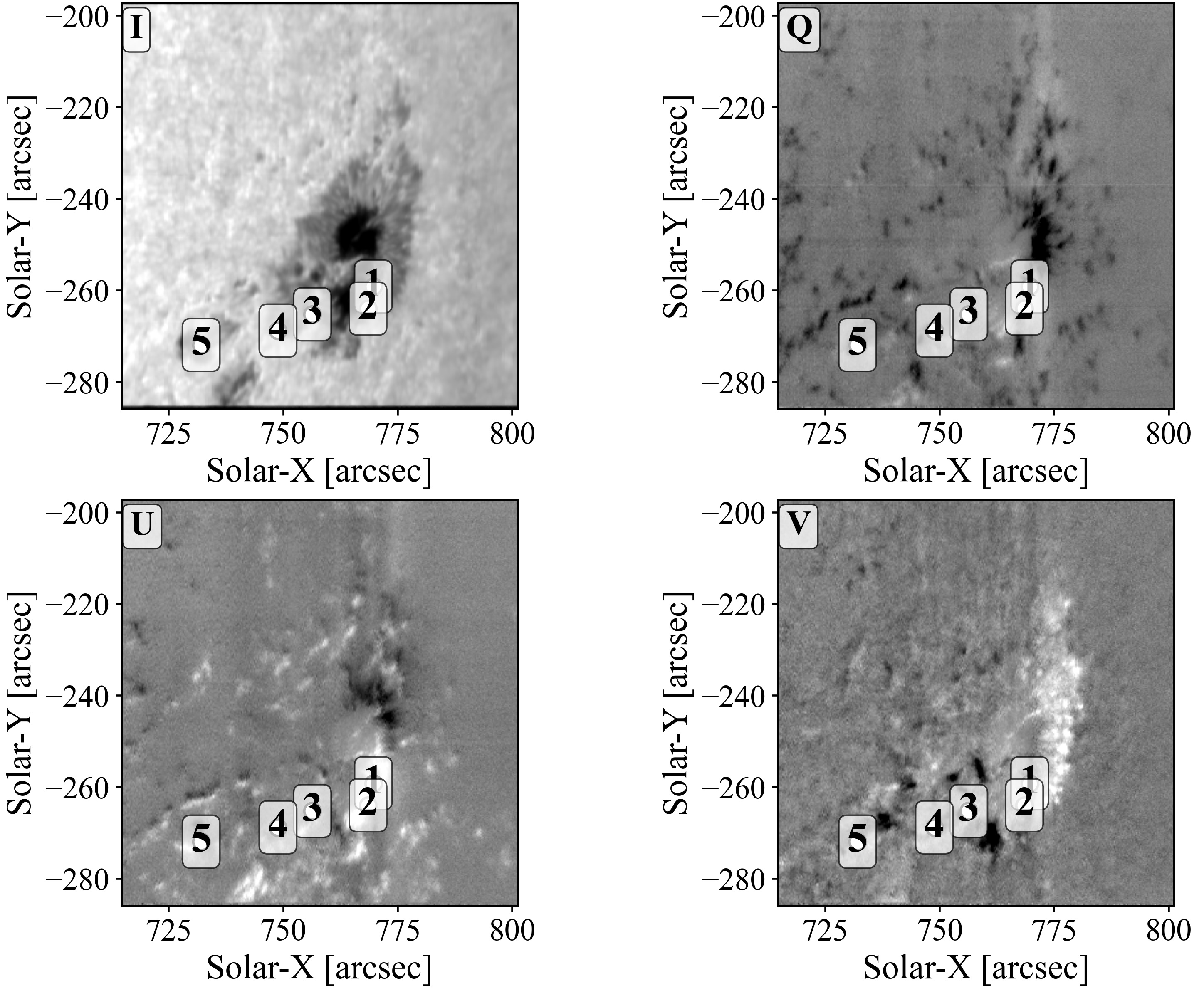}
    \subcaption{Core–band mean Stokes maps ($I,Q,U,V$; grayscale). 
    The gray box marks the quiet–Sun patch; numbered markers indicate foot–point regions (FPs).}
    \label{fig:tri-a}
  \end{subfigure}

  \vspace{0.6em}

  \begin{minipage}[t]{0.35\textwidth}
    \centering
    \includegraphics[width=\linewidth]{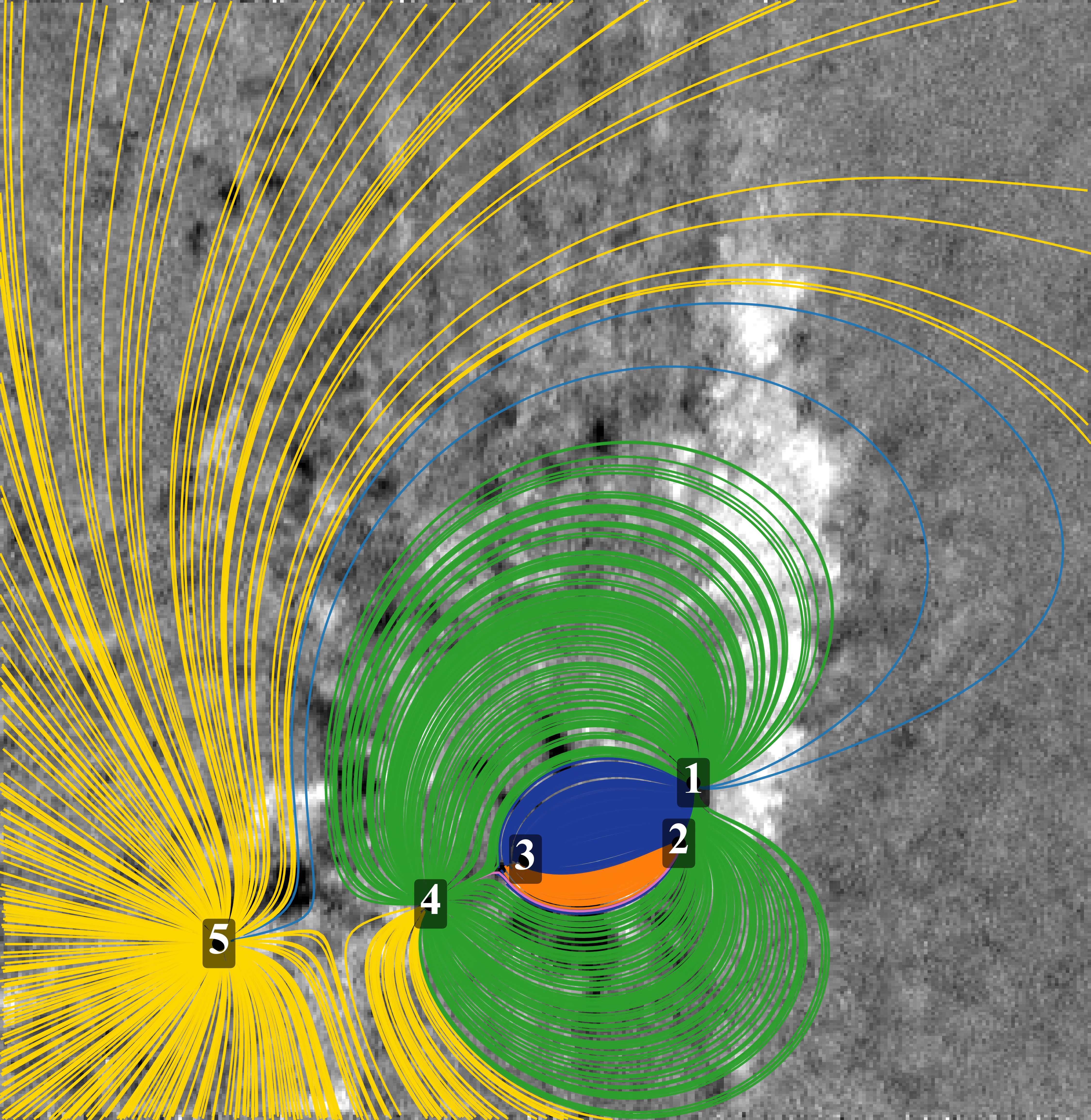}
    \subcaption{Pre–flare connectivity traced on the background image. 
    Streamlines from seed FPs are shown; trajectories that do not terminate at a sink FP are drawn as open.}
    \label{fig:tri-b}
  \end{minipage}
  \hspace{0.6 cm}
  \begin{minipage}[t]{0.35\textwidth}
    \centering
    \includegraphics[width=\linewidth]{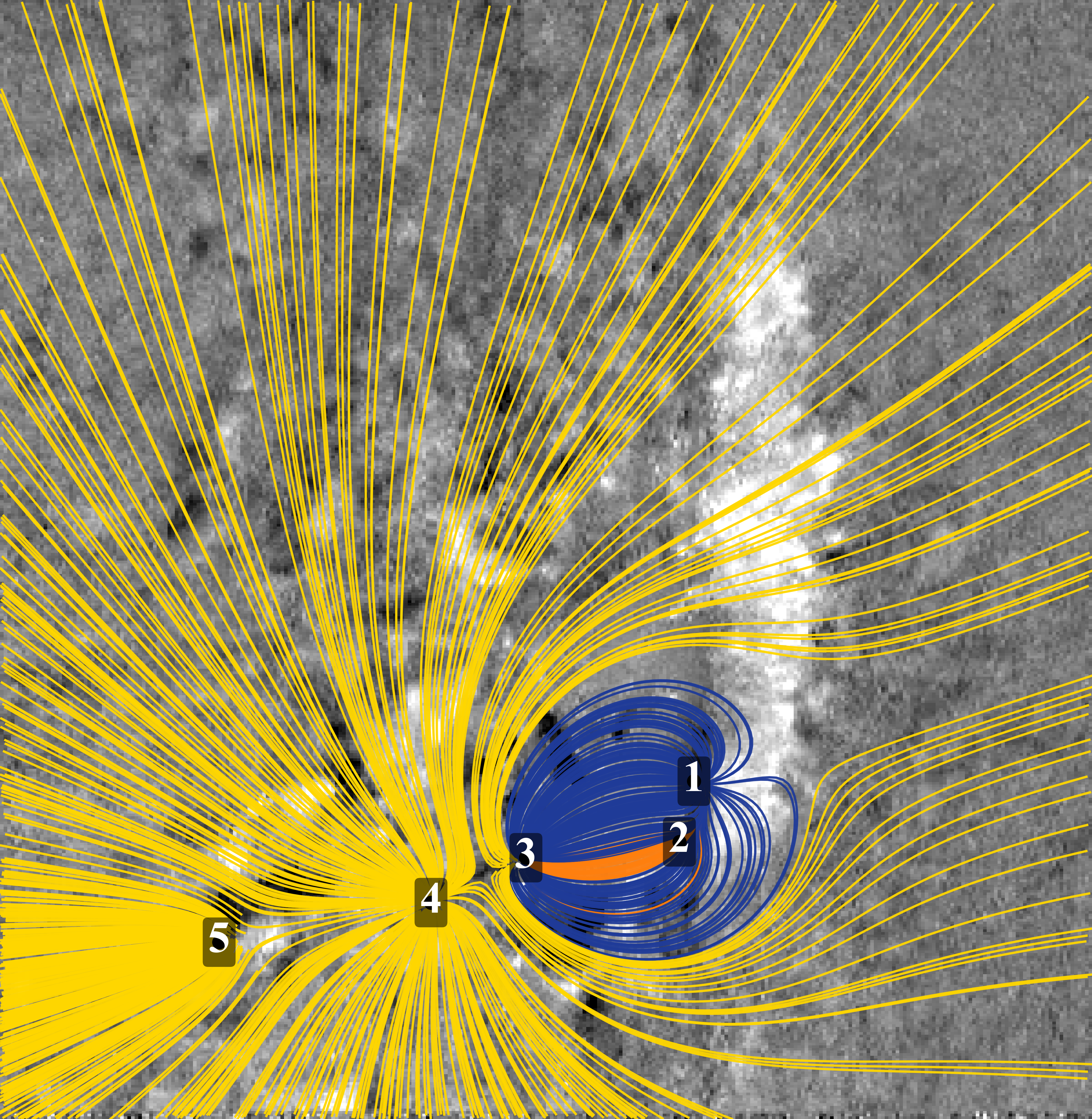}
    \subcaption{Post–flare connectivity with routing constraints (allow–map, quotas, gate). 
    Open trajectories use a single color; captured paths are colored by source$–$sink pair (color online).}
    \label{fig:tri-c}
  \end{minipage}

  \caption{Three–panel summary representing a sample magnetic configuration around the two components of flare trigger. 
   {(a)} Stokes $I,Q,U,V$ core–band means from \textsc{SPINOR}/DST (Fe\,\textsc{i}\,$\lambda$5250.2\,\AA); helioprojective axes are in arcsec, the quiet–Sun box is used for reference, and numbered markers denote the foot–point regions (FPs) used in the connectivity analysis. 
   {(b)} Pre–flare connectivity. 
   {(c)} Post–flare connectivity with the same seeds and sinks but with routing constraints.}
  \label{fig:tri-panel}
\end{figure*}

Figure~\ref{fig:tri-a} shows spectropolarimetric observations of the Fe~\textsc{i}~$\lambda5250.2$~\AA\ line. Observations were acquired with a spectrograph resolution, $R\approx145,000$ for 300 slit positions that densely scanned the main sunspot and surrounding areas with a step size of $0\farcs29$. The images are created from the bisector method, where we defined a “core band” around the line center using the paired–flank lobe-bisector procedure: for each Zeeman lobe, we specified a blue and a red window and a lobe core wavelength $\lambda_0$; for a set of fractional levels $f\in\{0.20,0.35,0.50,0.65,0.80,0.90\}$. An intensity target, $I_t = I_{\min} + f\,[\min(I_{\mathrm{edge,\,L}},I_{\mathrm{edge,\,R}})-I_{\min}]$ was computed, 
and solved on  {both} flanks for the two wavelengths $\lambda_{\mathrm{L}}(f)$ and $\lambda_{\mathrm{R}}(f)$ that satisfy $I(\lambda)=I_t$. 
Each cube was interpolated to all paired wavelengths and averaged to form a single "core-band mean" image per Stokes parameter (top panel in Figure ~\ref{fig:tri-a}); the gray rectangle outlines the quiet-Sun (QS) patch used for normalization/quality checks. 
 {This paired–flank lobe-bisector ``core-band'' definition is a straightforward extension of standard line-bisector techniques used in solar spectroscopy (e.g., \citealt{Deming2024}), rather than a completely new inversion method.}
The resulting $Q$, $U$, and $V$ maps reveal intricate polarization structure at the penumbral edge of the jet footpoints (FPs~1–2), flare footpoints (FPs 3 and 4) and the footpoints of flare ribbons (FP 5),  with $V$ showing small opposite-polarity kernels and $Q/U$ highlighting azimuthal variations of the transverse field; these features are co-spatial with the bright photospheric context seen in $I$.


Taken together, the $Q,U,V$ morphology points to multiple opposite polarities in close proximity and strong horizontal–field gradients at the penumbral boundary, consistent with stressed, non–potential fields where the jet and flare are anchored.  {We traced two-dimensional photospheric connectivity on the WCS grid using a point-source potential proxy field constructed from the flux centroids of identified foot-point (FP) markers. This follows the magnetic charge topology (MCT) formalism described by \citet{Longcope2005LRSP}, \citet{BarnesLongcopeLeka2005}, and \citet{TitovHornigDemoulin2002}. Each foot point~$i$ is represented by a point charge with signed weight $q_i$ at position $\boldsymbol{r}_i=(x_i,y_i)$, where the sign reflects its magnetic polarity. The resulting two-dimensional vector field is the superposition
\begin{equation}
  \boldsymbol{B}(\boldsymbol{r})
  = \sum_i q_i \,
    \frac{\boldsymbol{r}-\boldsymbol{r}_i}
         {|\boldsymbol{r}-\boldsymbol{r}_i|^2 + s^2},
\end{equation}
where the softening length $1\farcs6$ prevents singularities near the source points. This simplified potential field serves as a geometric guide for tracing streamline connectivity between opposite-polarity foot points on the photosphere. These observational patterns motivate the connectivity between the footpoints of the brightenings and the flaring regions summarized in Figure~\ref{fig:tri-panel}b,c.\ The streamline–tracing algorithm and quantitative connectivity statistics are described in detail in Appendix~\ref{app:magnetic_connection}.}

\subsection{H$\alpha$ fibrils connecting brightening to flares.}

H$\alpha$ fibrils in Figure~\ref{fig:bright_flare_goes} provide a chromospheric view of magnetic connectivity: in low-$\beta$ conditions they tend to align with the magnetic field and can be used as directional proxies for field lines, though notable departures are documented \citep{Leenaarts2012,DeLaCruz2011,AsensioRamos2017,Schad2013}. Flare ribbons observed in H$\alpha$ trace the chromospheric footprints of coronal reconnection at separatrices or quasi-separatrix layers (QSLs), and their morphology (e.g., two-ribbon, circular, and hook-shaped forms) reflects the underlying topology \citep{Fletcher2011,Masson2009FanSpine}.

In our event, bright fibrillar strands coalesce into a proxy ribbon system that straddles the polarity inversion line and maps the same footpoint linkages inferred from our \textsc{SPINOR} connectivity analysis, including transient brightenings near remote footpoints. The apparent ribbon separation and the unsigned magnetic flux swept by the ribbons provide standard estimates of the reconnection rate and its evolution during the flare \citep{Qiu2002,Asai2004}. Together, these H$\alpha$ diagnostics connect chromospheric morphology with changing magnetic connectivity, supporting a reconnection-driven reconfiguration scenario consistent with the topology indicated by our proxy connectivity maps.

\section{Concluding Remarks} \label{sec:conclusion}

Our pilot study shows that compact Ca\,\textsc{ii}  K brightenings located near the eventual flare kernel provide a repeatable early signature of major chromospheric activity. Across eight events, the precursor region peaks $\sim$11--46\,min before the primary Ca\,\textsc{ii}  K kernel at separations of $\sim$20--110\arcsec, and typically precedes the GOES 1--8\,\AA\ soft X-ray rise.  {When we fit a straight line to the ensemble
$\Delta t$--$D$ relation (Appendix~\ref{sec:appendixB}), the slope
corresponds to an apparent propagation speed of order 30--35\,km\,s$^{-1}$,
but the underlying correlations are weak and depend on linear/monotonic
statistics, so this value should be interpreted as an order-of-magnitude
indicator rather than a precise measurement.}

 {A second key result is that this behavior can be recovered using a deliberately simple analysis: fixed detector-coordinate ROIs, minimal preprocessing, and purely timing-based peak finding. By avoiding image segmentation, tracking, or heavy modeling, the measured offsets primarily reflect genuine solar evolution rather than registration artifacts. In this configuration, a compact Ca\,\textsc{ii} K brightening near a target kernel provides tens of minutes of actionable lead time, while the subsequent kernel peak marks the imminent soft X-ray maximum.  {However, in some events the Ca\,\textsc{ii}~K emission near the flare site shows multiple peaks in several ROIs (Section~\ref{sec:appendixA}), so the brightening should at present be regarded as a promising but not yet standalone flare-onset proxy in complex or highly variable active regions.} Methodological caveats (e.g., partial temporal coverage, closely spaced sub-flares, and minute-quantized GOES times) are discussed in Section~\ref{sec:methodsandresult}, but they do not erase the multi-minute lead times seen across the ensemble.}

 {Our \textsc{SPINOR} example adds a complementary, magnetic perspective. A simple point-source proxy field, constructed from footpoint centroids, reveals that pre-flare connectivity is dominated by compact closed links (notably FP$_3\!\rightarrow\!$FP$_2$ and FP$_4\!\rightarrow\!$FP$_1$), whereas post-flare connectivity is rerouted toward more open or extended trajectories, especially from FP$_3$ and FP$_5$. Together with the H$\alpha$ fibrillar and ribbon morphology, this pattern is consistent with reconnection that removes low-lying closed flux and transfers it to more extended connectivities in breakout, tether-cutting, or fan–spine–like scenarios. Even though the proxy field is not a full PFSS/NLFFF extrapolation, it provides a reproducible way to track topological changes with minimal assumptions.}

 {Looking ahead, the immediate priorities are to expand the Ca\,\textsc{ii}  K sample, quantify forecast skill (lead-time distributions and false-alarm rates), and incorporate contextual gating such as plage masks and ribbon-based constraints. Automated ROI placement and flux-weighted connectivity statistics from vector inversions will allow us to connect topology, energetics, and precursor behavior more directly. In combination, these developments can turn the simple timing signature identified here into a deployable, ground-based flare nowcasting module that is both operationally lightweight and physically interpretable.}

\begin{acknowledgments}
AK, SS, and JS are funded by NSF grants 1936336, 2401175, and NASA's grant NNH23ZDA001N-BPSF. DC acknowledges partial support for this project from NASA grants 19-HSODS-004 and 21-SMDSS21-0047. We thank DST observers Colin Hancock, Sara Jefferies, and Shane Thompson, who acquired data used for this research.

\end{acknowledgments}

\appendix
\restartappendixnumbering

\section{Appendix A: Supplementary Methods and Consistency Checks}
\label{sec:appendixA}

This appendix expands on several aspects of the analysis, including the light-curve detrending and peak-identification procedure, a note on false-positive candidates, a check on the geometric foreshortening correction, and the complete per-event set of Ca\,\textsc{ii}\,K light curves used to measure the temporal offsets ($\Delta t$) between compact brightenings and flare kernels.

\paragraph{Detrending and Peak Identification.}
For every region of interest (ROI), we constructed a mean-intensity light curve at the native cadence and removed slow intensity trends by fitting and subtracting a low-order (quadratic, by default) polynomial in time. Peaks in the detrended series $\Delta I(t)$ were then identified using a prominence threshold greater than three times the running-median background level and by locating zero crossings in the first derivative.  
To suppress frame-to-frame jitter while maintaining native timing precision, we applied a light smoothing over $w\in\{1,3,5\}$ frames and recorded the vertex time from a quadratic fit within a narrow window ($\pm$3–5 frames) around the detected maximum.

\begin{figure*}[ht!]
  \centering
  \includegraphics[width=0.8\textwidth]{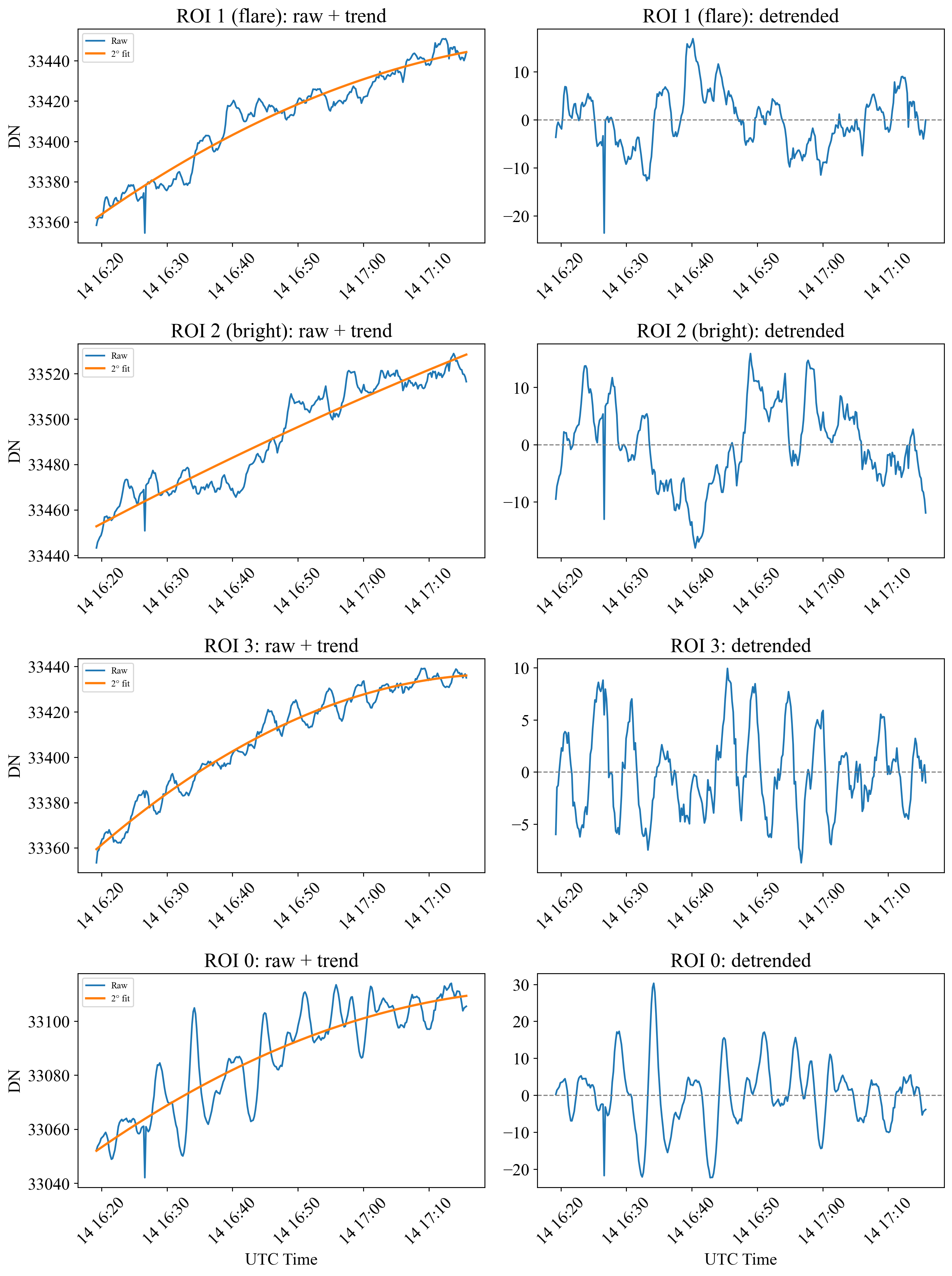}
  \caption{Example of detrending and ambiguity in precursor identification for the 2021-09-14 event. Left column: raw Ca\,\textsc{ii}\,K light curves with quadratic background fits (orange). Right column: detrended intensity series $\Delta I$ used for timing analysis. Several ROIs brighten before the main flare kernel, producing potential false positives that are currently resolved by visual inspection.}
  \label{fig:app:detrend}
\end{figure*}

\paragraph{Visual Screening and False Positives.}
Figure~\ref{fig:app:detrend} illustrates the detrending procedure for the 2021-09-14 event and highlights an important limitation of purely intensity-based selection. In several cases, more than one ROI exhibits an impulsive rise prior to the main flare kernel (for example, Regions~4 and~5), producing multiple potential “precursor” candidates.  
At present there is no fully automated way to determine which brightening is magnetically or physically linked to the subsequent flare. This ambiguity is mitigated by visual inspection of the Ca\,\textsc{ii}\,K and H$\alpha$ context images and by temporal cross-checking, but a robust, machine-based disambiguation—perhaps using morphological or magnetic connectivity constraints—remains a goal for future work.

\begin{figure}[h!]
  \centering
  \includegraphics[width=0.56\textwidth]{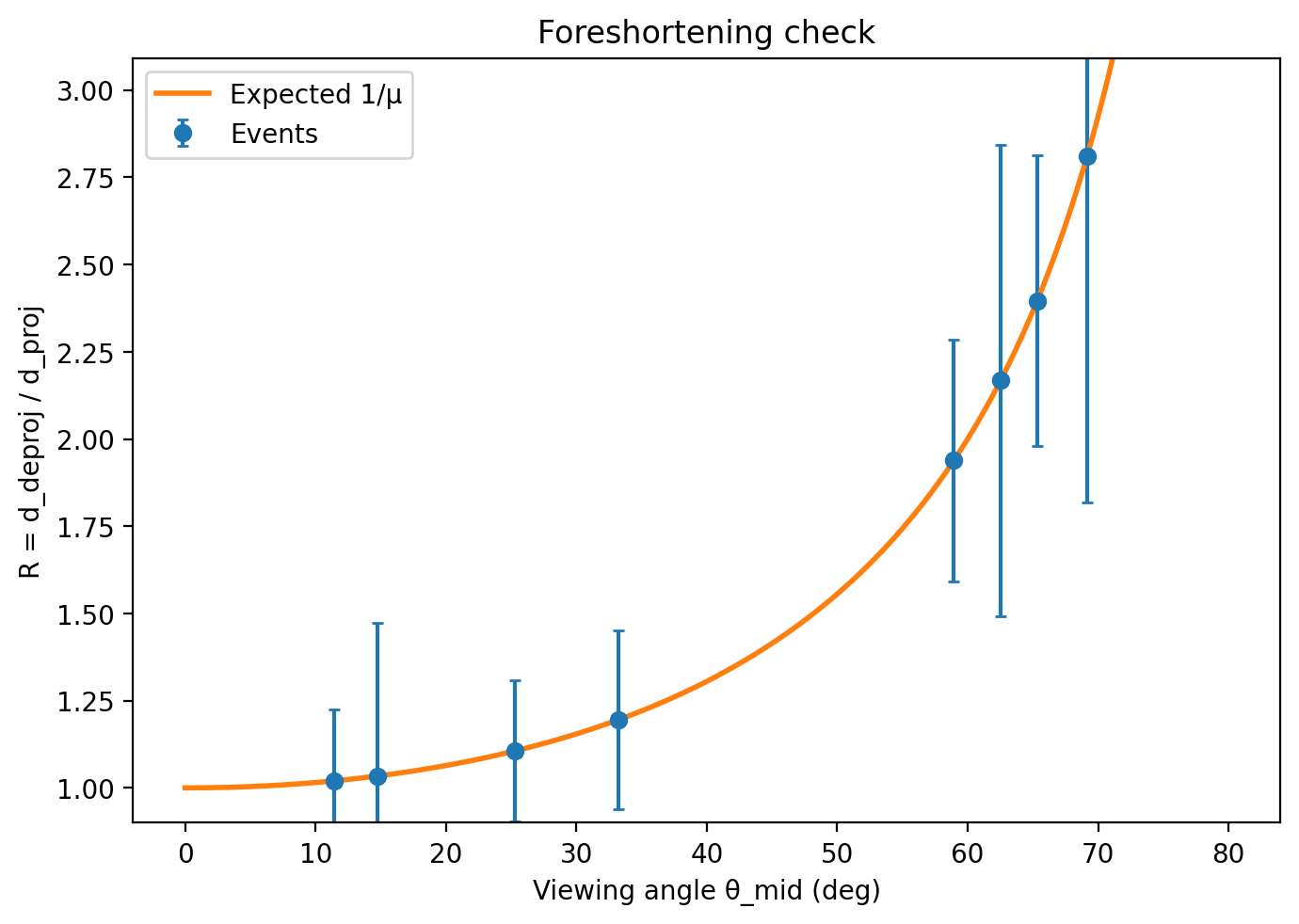}
  \caption{Foreshortening consistency check. Event ratios $R=d_{\rm deproj}/d_{\rm proj}$ (blue symbols, $1\sigma$ error bars) are plotted against the heliocentric angle $\theta_{\rm mid}$ and compared with the geometric expectation $1/\mu$ (orange curve), where $\mu=\cos\theta_{\rm mid}$. Error construction is described in Appendix~\ref{sec:app:errors}.}
  \label{fig:app:foreshortening}
\end{figure}

\paragraph{Foreshortening Consistency.}
We tested the accuracy of our geometric deprojection by comparing the measured deprojected-to-projected distance ratio, $R=d_{\rm deproj}/d_{\rm proj}$, for each event against the simple geometric expectation $1/\mu$ with $\mu=\cos\theta_{\rm mid}$, where $\theta_{\rm mid}$ is the heliocentric angle at the midpoint of the brightening–flare pair. The strong agreement between the measured ratios and $1/\mu$, shown in Figure~\ref{fig:app:foreshortening}, confirms that the applied deprojection correctly accounts for the viewing geometry and that the WCS headers provide consistent disk positioning.

\begin{figure*}[h!]
  \centering
  \includegraphics[width=\textwidth]{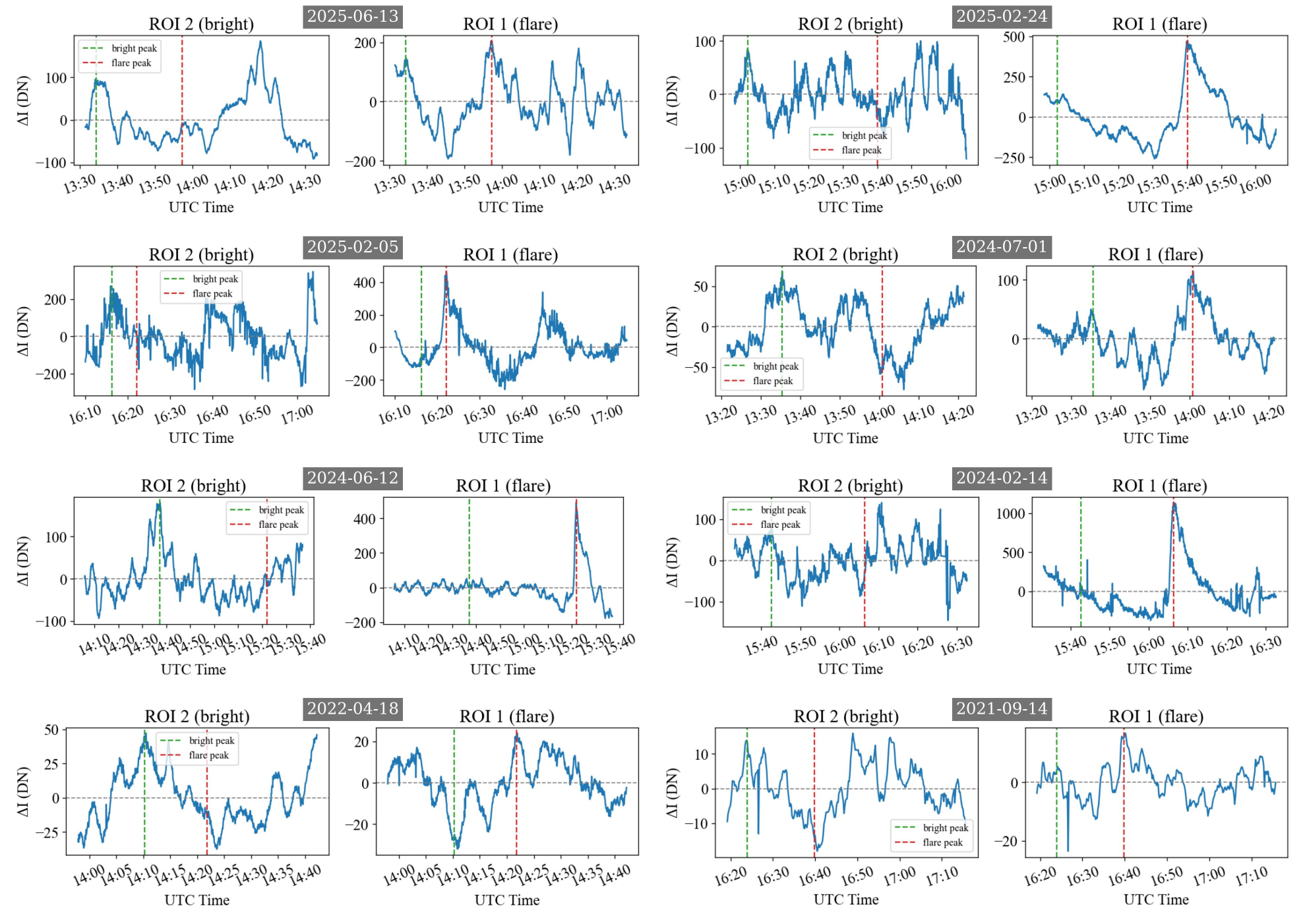}
  \caption{Detrended Ca\,\textsc{ii}\,K light curves for all eight events. For each date, the compact brightening ROI (left) and flare kernel (right) are shown with dashed green and red lines marking their respective peak times. These curves are the direct notebook outputs used to determine the temporal offsets $\Delta t$.  {ROI~1 is the flare kernel and ROI~2 the precursor brightening for every event, while ROI~0 (when present) is a quiescent reference region; other ROIs are additional trial locations that do not show significant impulsive peaks.}}
  \label{fig:app:all8}
\end{figure*}

\paragraph{Per-Event Light Curves.}
For completeness, Figure~\ref{fig:app:all8} presents the detrended Ca\,\textsc{ii}\,K light curves for all eight analyzed events. Each panel pair shows the compact brightening ROI (left) and the corresponding flare kernel (right), with dashed green and red lines marking their respective peak times. These plots are the direct outputs from the analysis notebooks and form the basis of the measured $\Delta t$ values used in the timing regressions presented in the main text.

\section{Appendix B: Distance--time correlations}
\label{sec:appendixB}

\begin{figure*}[ht!]
\centering
\includegraphics[width=0.75\linewidth]{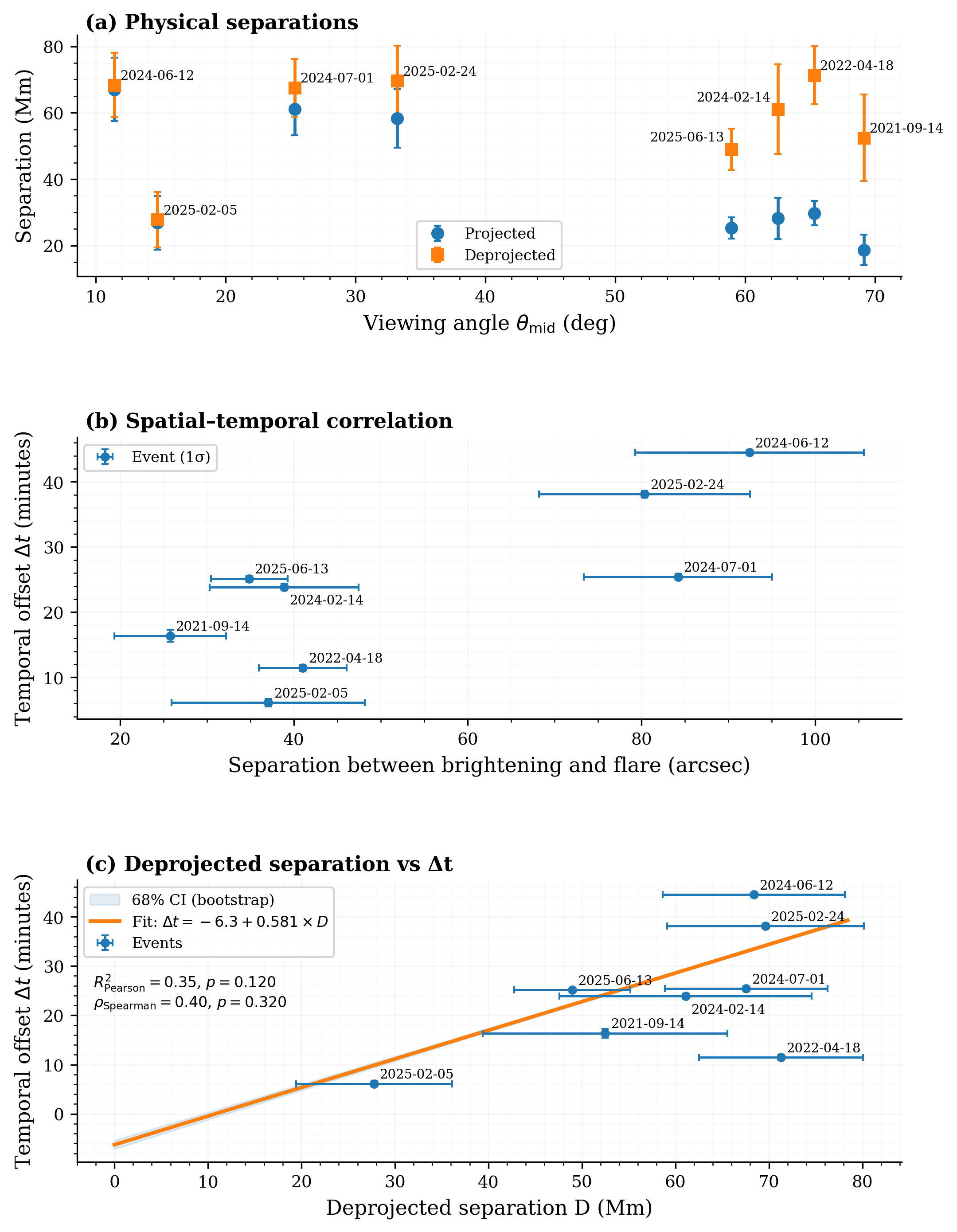}
\caption{
Relationships between viewing geometry, spatial separation, and timing of precursor--flare pairs.
(a) Projected (blue circles) and deprojected (orange squares) separations between the precursor brightening and the primary flare kernel as a function of heliocentric viewing angle $\theta_{\rm mid}$, with 1$\sigma$ errors.
(b) Temporal offsets ($\Delta t$) versus projected separations in arcseconds.
(c) $\Delta t$ as a function of deprojected distance $D$, with best-fit linear trend (orange) and 68\% bootstrap confidence interval (shaded). 
The correlation coefficients shown refer to Pearson's $R^2$ and Spearman's $\rho$.  {Because these metrics assume linear or monotonic relationships and the coefficients are only modest, the $\Delta t$--$D$ trend should be regarded as suggestive rather than definitive; the implied apparent speed (30--35\,km\,s$^{-1}$) is therefore an order-of-magnitude indicator rather than a precise measurement.}}
\label{fig:geom_timing_speed}
\end{figure*}

 {In this appendix we summarize the ensemble behavior of the precursor--flare timing offsets as a function of spatial separation. Figure~\ref{fig:geom_timing_speed} compiles the projected and deprojected distances between the precursor and flare kernels, together with the corresponding temporal offsets $\Delta t$. A weighted least-squares fit to $\Delta t$ versus $D$ yields a slope equivalent to an apparent propagation speed of order 30--35\,km\,s$^{-1}$, but the Pearson and Spearman correlation coefficients are only modest ($\lvert r\rvert \lesssim 0.5$) and assume linear or monotonic relationships. The distance--time trend should therefore be viewed as a weak, illustrative correlation rather than a strong predictive law, and the inferred speed is best interpreted as an order-of-magnitude guide.}

\section{Uncertainties and Error Propagation}
\label{sec:app:errors}

This section details how the timing and geometric uncertainties were estimated and propagated into the error bars shown in the figures and regressions.  
In what follows, $\delta t$ denotes the native cadence, $\Delta t$ is the measured brightening–flare peak offset, $d_{\rm proj}$ is the projected on-disk distance between ROI centroids, $\theta_{\rm mid}$ is the heliocentric angle at the midpoint of the pair, $\mu=\cos\theta_{\rm mid}$, and $D=d_{\rm proj}/\mu$ is the corresponding deprojected distance.

\subsection{Timing Uncertainties}

The total timing uncertainty for the peak of a single ROI light curve, $\sigma_t$, is computed as the quadrature sum of three independent contributions:
\begin{equation}
  \sigma_t^2
  \;=\;
  \underbrace{\frac{\delta t^2}{12}}_{\text{sampling}} \;+\;
  \underbrace{\sigma_{t,{\rm fit}}^2}_{\text{local quadratic fit}} \;+\;
  \underbrace{\sigma_{t,{\rm meth}}^2}_{\text{method robustness}} .
\end{equation}

The first term represents the sampling error and assumes that the true peak can occur anywhere within a single frame.  
The second term, $\sigma_{t,{\rm fit}}$, is obtained from the covariance matrix of a quadratic fit performed over a $\pm N$-frame window (typically $N=3$–5) centered on the detected maximum, and represents the local vertex uncertainty.  
The third term quantifies the sensitivity of the peak time to small changes in the smoothing or detrending procedure.  
It is defined as the median absolute deviation (MAD) of the peak times recomputed across smoothing windows $w\in\{1,3,5\}$, combined with the MAD obtained by varying the detrending polynomial degree between one and three:
\begin{equation}
  \sigma_{t,{\rm meth}}^2
  \;=\;
  \mathrm{MAD}_w^2\big(t_{\rm peak}\big)
  \;+\;
  \mathrm{MAD}_{\rm deg}^2\big(t_{\rm peak}\big).
\end{equation}

The uncertainty in the time difference between the brightening and flare peaks, $\Delta t$, is then given by
\begin{equation}
  \sigma_{\Delta t}
  \;=\;
  \sqrt{
    \sigma_{t,{\rm bright}}^2
    \;+\;
    \sigma_{t,{\rm flare}}^2
    \;+\;
    \sigma_{t,{\rm GOES}}^2
  } ,
\end{equation}
where $\sigma_{t,{\rm GOES}}=\delta t_{\rm GOES}/\sqrt{12}$ accounts for the one-minute sampling of the GOES 1–8\,\AA\ light curves and is included only when a GOES timestamp directly enters the timing offset.

\subsection{Projected and Deprojected Separations}

The projected separation between two ROIs, $d_{\rm proj}$, is computed from their centroid coordinates $(x_1,y_1)$ and $(x_2,y_2)$ on the solar disk as
\begin{equation}
  d_{\rm proj}=\sqrt{(x_2-x_1)^2+(y_2-y_1)^2}.
\end{equation}
Its uncertainty depends on the measurement precision of those centroids and the accuracy of the plate scale.  
The propagated variance is
\begin{equation}
  \sigma_{d_{\rm proj}}^2 =
  \left(\frac{\Delta x}{d_{\rm proj}}\right)^2(\sigma_{x_1}^2+\sigma_{x_2}^2)
  +
  \left(\frac{\Delta y}{d_{\rm proj}}\right)^2(\sigma_{y_1}^2+\sigma_{y_2}^2),
\end{equation}
where $\Delta x=x_2-x_1$ and $\Delta y=y_2-y_1$.  
We adopt a conservative centroid error of $0.3$–$0.5$\,pix per coordinate and include an additional fractional plate-scale term $\epsilon_{\rm ps}$ when available, such that
\begin{equation}
  \sigma_{d_{\rm proj}}^2
  \;\rightarrow\;
  \sigma_{d_{\rm proj}}^2
  + (\epsilon_{\rm ps}\,d_{\rm proj})^2 .
\end{equation}

The deprojected separation $D=d_{\rm proj}/\mu$ accounts for the foreshortening of features observed away from disk center.  
The uncertainty in $\mu=\cos\theta_{\rm mid}$ follows from propagation of the angular uncertainty $\sigma_\theta$ as
\begin{equation}
  \sigma_{\mu} = |\sin\theta_{\rm mid}|\,\sigma_{\theta}.
\end{equation}
The corresponding uncertainty in $D$ is therefore
\begin{equation}
  \sigma_{D}^2
  =
  \frac{\sigma_{d_{\rm proj}}^2}{\mu^2}
  +
  \left(\frac{d_{\rm proj}}{\mu^2}\right)^2
  \sigma_{\mu}^2 .
\end{equation}

When examining the foreshortening ratio $R\equiv D/d_{\rm proj}$, we use the relation $R=1/\mu$, for which the uncertainty becomes
\begin{equation}
  \sigma_R = \frac{\sigma_\mu}{\mu^2}.
\end{equation}
This expression avoids introducing correlated terms between $D$ and $d_{\rm proj}$ and provides a direct geometric comparison with the expectation from helioprojective coordinates.

\subsection{Apparent Propagation Speed}

The apparent propagation speed is defined as the ratio of deprojected separation to timing offset,
\begin{equation}
  v = \frac{D}{\Delta t}.
\end{equation}
The corresponding uncertainty is obtained through standard error propagation as
\begin{equation}
  \sigma_v
  =
  v \,
  \sqrt{
    \left(\frac{\sigma_D}{D}\right)^2
    +
    \left(\frac{\sigma_{\Delta t}}{\Delta t}\right)^2
  } .
\end{equation}
This quantity represents the effective speed at which the disturbance associated with the compact brightening needs to travel to account for the measured spatial and temporal offsets.

 {\section{Magnetic connection}\label{app:magnetic_connection}}
 {In this appendix we complement the qualitative SPINOR connectivity maps in Section~\ref{sec:spinor_maps} with a simple, streamline–based measure of how the photospheric linkages between the marked footpoints (FP$_1$–FP$_5$) change across the flare. Using the point–source proxy field constructed from the flux–weighted centroids of these regions, we treat FP$_3$–FP$_5$ as source footpoints and FP$_1$, FP$_2$ as candidate sinks, and then follow large ensembles of field–line proxies in the helioprojective plane. The goal is not to recover a full coronal extrapolation, but to quantify in a reproducible way the fraction of trajectories that remain in short, closed connections versus those that are redirected into more extended or open configurations before and after the flare, thereby turning the topology suggested by Figure~\ref{fig:tri-panel} into simple connectivity statistics.}

 {To quantify connectivity we launched $N=250$ streamline seeds from each of FP$_3$, FP$_4$, and FP$_5$ and integrated the unit proxy field $\hat{\boldsymbol b}=(B_x,B_y)/\!\sqrt{B_x^2+B_y^2}$ built from signed point–fluxes at the FP markers, as introduced in Section~\ref{sec:spinor_maps}. Streamlines were integrated in the helioprojective plane with a fourth–order Runge–Kutta scheme (step $0\farcs55$, maximum 7000 steps); a trajectory is deemed \emph{captured} by a sink FP (here FP$_1$, FP$_2$) when it comes within $2^{\prime\prime}$ of that FP, otherwise it is classified as \emph{open}. The Stokes layer shown underneath is only a background for context; the trajectories are traced in the proxy field, not in the image layer. We place stochastic seeds (250 per FP) within a $\sim0\farcs9$ Gaussian around each source FP and integrate streamlines of the two-dimensional proxy field $\boldsymbol{B}(x,y)$ introduced in Section~\ref{sec:spinor_maps}. A path is marked ``captured'' if it approaches a sink FP within $2\arcsec$; otherwise it is labeled ``open.''}

 {The pre–flare conditions show: from FP$_3$, $29/250=11.6\%$ of streamlines terminate at FP$_1$ and $221/250=88.4\%$ at FP$_2$; from FP$_4$, $184/250=73.6\%$ connect to FP$_1$ and $24/250=9.6\%$ to FP$_2$ (the remaining $16.8\%$ do not capture within the domain); from FP$_5$, $23/250=9.2\%$ reach FP$_1$ and $0/250=0.0\%$ reach FP$_2$ (the remaining $90.8\%$ are non–capturing). The post–flare solution with routing constraints shows that nearly all trajectories from FP$_3$ are open ($247/250=98.8\%$), FP$_5$ is likewise predominantly open ($230/250=92.0\%$), while FP$_4$ shows $101/250=40.4\%$ open and the remainder captured by one of the allowed sinks.}

 {Before the flare the field is largely closed between the FP pairs: FP$_3\!\rightarrow\!$FP$_2$ (about 88\%) and FP$_4\!\rightarrow\!$FP$_1$ (about 74\%), whereas FP$_5$ is mostly non–capturing (about 91\% open). After the flare the pattern reorganizes dramatically: trajectories from FP$_3$ become almost entirely open ($\sim$99\%), FP$_5$ remains predominantly open ($\sim$92\%), and FP$_4$ shows a substantial increase in openness ($\sim$40\% open) with the remaining paths closing mainly to FP$_1$. In other words, the post–flare state redirects much of the FP$_3$ and FP$_5$ connectivity from compact, closed links into large–scale/open connections, while FP$_4$ retains only a minority of closed flux to FP$_1$. This point–source field is a coarse, 2-D proxy meant to illustrate photospheric connectivity. It is not a replacement for coronal field extrapolations (PFSS/NLFFF) or for a full magnetic–charge topology (MCT) analysis based on partitioned magnetograms; those methods typically use many sources from magnetograms, whereas we use only a few ROI footpoints \citep{Longcope2005LRSP,WiegelmannSakurai2012,BarnesLongcopeLeka2005}.}

 {The connectivity changes we see, involves loss of short, closed flux near penumbral footpoints and growth of longer connections, are consistent with reconnection that removes low–lying flux and transfers it to more extended connectivities, as in breakout or tether–cutting scenarios and in fan–spine/circular–ribbon events \citep{Antiochos1999,Moore2001,Masson2009FanSpine}. A delayed, multi–stage reconnection sequence involving a three–ribbon/fan–spine topology is therefore plausible and will be tested in follow–up work \citep{Sun2013HotSpine,Dudik2014Slipping,Li2016Xshape}.}


\bibliography{main_paper}{}



\end{document}